\documentstyle[pra,multicol,aps,epsfig]{revtex}
\begin{document}
\renewcommand{\textfraction}{0.10} 
\renewcommand{\topfraction}{1.0} 
\renewcommand{\bottomfraction}{1.0} 
\flushbottom
\title{Dynamics of Atom-Mediated Photon-Photon Scattering II: 
Experiment}
\author{M. W. Mitchell, Cindy I. Hancox and R. Y. Chiao}
\address{Department of Physics, University of California at Berkeley,
Berkeley, CA 94720, USA  }

\date{\today} 
\maketitle

\begin{abstract}
    Temporal and angular correlations in atom-mediated photon-photon 
    scattering are measured.  Good agreement is found with the theory 
    presented in Part~I. 
\end{abstract}


\begin{multicols}{2}
\section{Introduction}

As described in Part I \cite{Mitchell=1999}, atom-mediated 
photon-photon scattering is the microscopic process underlying the 
optical Kerr nonlinearity in atomic media.  The Kerr nonlinearity 
produces such effects as self-phase modulation, self-focusing and 
self-defocusing and four-wave mixing.  In an atomic medium, resonanant 
nonlinearities can give rise to very large nonlinear optical effects, 
suggesting the possibility of nonlinear optical interactions with only 
a few photons \cite{Imamoglu=1997}.  Unfortunately, the same 
resonances which could facilitate such experiments make them difficult 
to analyze \cite{Grangier=1998}.  In Part I we showed theoretically 
that the photon-photon interaction is not intrinsically lossy, and can 
be fast on the time scale of atomic relaxation.  Here we describe an 
experiment to directly measure the time-duration of the photon-photon 
interaction in a transparent medium.

In the scattering experiment, two off-resonance laser beams collide in 
a rubidium vapor cell and scattering products are detected at right 
angles.  The process of phase-matched resonance fluorescence in this 
geometry has been described as spontaneous four-wave mixing 
\cite{Heidmann=1987}, a description which applies to our off-resonant 
excitation as well.  This geometry has been of interest in quantum 
optics for generating phase-conjugate reflection \cite{Gaeta=1988}. 
Elegant experiments 
with a barium atomic beam \cite{Grangier=1986} showed antibunching in 
multi-atom resonance fluorescence, but a separation of timescales was 
not possible since the detuning, linewidth and doppler width were all 
of comparable magnitude.

\section{Setup}

A free-running 30mW diode laser at 780nm was temperature stabilized 
and actively locked to the point of minimum fluorescence between the 
hyperfine-split resonances of the D2 line of rubidium.  Saturation 
spectroscopy features could be observed using this laser, indicating a 
linewidth $\delta\nu < 200$ MHz.  This linewidth is small compared 
with the detuning from the nearest absorption line $\delta \nu= 1.3$ 
GHz .  Direct observation of the laser output with a fast photodiode 
(3 dB rolloff at 9 GHz) showed no significant modulation in the 
frequency band $100$ MHz -- $2$ GHz.  The laser beam was was shaped by 
passage through a single-mode polarization-maintaining fiber, 
collimated and passed through a scattering cell to a retro-reflection 
mirror.  The beam within the cell was linearly polarized in the 
vertical direction.  The beam waist (at the retroreflection mirror) 
was $0.026$ cm $\times 0.023$ cm (intensity FWHM, vertical $\times$ 
horizontal).  The center of the cell was 1.9 cm from the 
retroreflection mirror, thus within a Rayleigh range of the waist.  
With optimal alignment, the laser could deliver 1.95 mW to the cell, 
giving a maximal Rabi frequency of $\Omega_{\rm Rabi} \approx 2 \times 
10^{9} $s$^{-1}$, significantly less than the minimal detuning of $ 
\delta = 2 \pi \times 1.3$ GHz $= 8 \times 10^{9} $s$^{-1}$.  For this 
reason, we have neglected saturation of the transitions in the 
analysis.

The retro-reflected beam returned through the fiber and was picked off 
by a beamsplitter.  The single-mode fiber acted as a near-ideal 
spatial filter and the returned power through the fiber provides a 
quantitative measure of the mode fidelity on passing through the 
rubidium cell.  With optimal alignment it was possible to achieve a 
mode fidelity (described below) of 36\%.

The cell, an evacuated cuvette filled with natural abundance rubidium 
vapor, was maintained at a temperature of 330 K to produce a density 
of about 1.6 $\times 10^{10}$ cm$^{-3}$.  Irises near the cell limited 
the field of view of the detectors.  Stray light reaching the 
detectors was negligible, as were the detectors' dark count rates of 
$< 100$ cps.

With the aide of an auxiliary laser beam, two single-photon counting 
modules (SPCMs) were positioned to detect photons leaving the 
detection region in opposite directions.  In particular, photons 
scattered at right-angles to the incident beams and in the direction 
perpendicular to the drive beam polarization were observed.  Each 
detector had a 500 $\mu$m diameter active area and a quantum 
efficiency of about 70\%.  The detectors were at a distance of 70 cm 
from the center of the cell.  The effective position of one detector 
could be scanned in two dimensions by displacing the alignment mirrors 
with inchworm motors.  A time-to-amplitude converter and multichannel 
analyzer were used to record the time-delay spectrum.  The system time 
response was measured using sub-picosecond pulses at 850nm as an 
impulse source.  The response was well described by a Lorentzian of 
width 810 ps (FWHM).

% \begin{figure}
% \begin{center}
% \epsfig{file=setup.eps,width = \figwidth ,angle=0}
% \mycaption{FIG. 1. Experiment schematic.}
% \label{setupFig}
% \end{center}
% \end{figure}
% 
% NOTE TO EDITOR: figure 1 goes near here.

Optimal alignment of the laser beam to the input fiber coupler could 
not always be maintained against thermal drifts in the laboratory.  
This affected the power of the drive beams in the cell but not their 
alignment or beam shape.  These were preserved by the mode-filtering 
of the fiber.  Since the shape of the correlation function depends on 
beam shape and laser tuning but not on beam power, this reduction in 
drive power reduced the data rate but did not introduce errors into 
the correlation signal.

\subsection{Experimental Results}
\index{correlation!time}

% \begin{figure}
% \begin{center}
% \epsfig{file=ExptTimes.eps,width = \figwidth,angle=0}
% \mycaption{FIG. 2.  Observed coincidence rates for right-angle photon-photon 
% scattering.  Circles show data acquired with detectors aligned to 
% collect back-to-back scattering products.  Squares show data acquired with 
% detectors misaligned by 10 cm $\approx$ 0.14 radian.  The solid line 
% is a Gaussian function fit to the data.  }
% \label{ExptTimesFig}
% \end{center}
% \end{figure}
% 
% NOTE TO EDITOR: figure 2 goes near here.
The time-delay spectrum of a data run of 45 hours is shown in Fig.  2.  
The detectors were placed to collect back-to-back scattering products 
to maximize the photon-photon scattering signal.  A Gaussian function 
$P(t_{A}-t_{B})$ fitted to the data has a contrast 
$[P(0)-P(\infty)]/P(\infty)$ of $0.046 \pm 0.008$, a FWHM of $1.3 \pm 
0.3$ ns, and a center of $-0.07 \pm 0.11$ ns.  This center position is 
consistent with zero, as one would expect by the symmetry of the 
scattering process.  For comparison, a reference spectrum is shown.  
This was taken under the same conditions but with one detector 
intentonally misaligned by much more than the angular width of the 
scattering signal.

% \begin{figure}
% \begin{center}
% \epsfig{file=angular2.eps,width = \figwidth,angle=0} 
% \mycaption{FIG. 3.  Signal contrast vs.  detector displacement.  A 
% displacement of 1 mm corresponds to an angular deviation of 1.43 mrad.  
% }
% \label{AngularFig}
% \end{center}
% \end{figure}
% NOTE TO EDITOR: figure 3 goes near here.

The angular dependence of the scattering signal was investigated by 
acquiring time-delay spectra as a function of detector position.  To 
avoid drifts over the week-long acquision, the detector was scanned in 
a raster pattern, remaining on each point for 300 s before shifting to 
the next.  Points were spaced at 1 mm intervals.  Total live 
acquisition time was 9 hours per point.  The aggregate time-spectrum 
from each location was fitted to a Gaussian function with fixed width 
and center determined from the data of Fig.  2.  The 
position-dependent contrast C(x,y) is shown in Fig.  3.  A negative 
value for the contrast means that the best fit had a coincidence 
dip rather than a coincidence peak at zero time.  These negative 
values are not statistically significant.  Fitted to a Gaussian 
function, C(x,y) has a peak of 0.044 $\pm .010$ and angular widths 
(FWHM) of $1.1 \pm 0.7$ mrad and $3.7 \pm 0.4$ mrad in the horizontal 
and vertical directions, respectively.

These angular widths are consistent with the expected coherence of 
scattering products \cite{Pittman=1995}.  Seen from the detector positions, the excitation 
beam is narrow in the vertical direction, with a Gaussian shape of 
beam waist $w_{y} = 0.009$ cm, but is limited in the horizontal 
direction only by the apertures, of size $\Delta z = 0.08 $ cm.  Thus 
we expect angular widths of 
$
%\Delta \theta_{\rm Horizontal} = 
0.9 {~\rm mrad}$ and 
$
%\Delta \theta_{\rm Vertical} = 
3.25 {~\rm mrad}$, 
where the first describes diffraction of a Gaussian, the second 
diffraction from a hard aperture.  

\section{Comparison to theory}

% \begin{figure}
% \begin{center}
% \epsfig{file=RbIdeal.eps01,width=\mathsize,angle=0}
% \mycaption{FIG. 4.  Coincidence rates by photon-photon scattering theory: 
% ideal case.}
% \label{PPIdealFig}
% \end{center}
% \end{figure}
% NOTE TO EDITOR: figure 4 goes near here.

The correlation signal predicted by the theory of Part I 
is shown in Fig.  4.  The ideal contrast is 1.53 and the FWHM is 870 
ps.  The shape of the time correlations is altered by experimental 
limitations.  First, beam distortion in passing through the cell 
windows reduces the photon-photon scattering signal.  Second, finite 
detector response time and finite detector size act to disperse the 
signal.  None of these effects alters the incoherent scattering 
background.

% \begin{figure}
% \begin{center}
% \epsfig{file=fidelity.eps01,width = \figwidth,angle=0}
% \mycaption{FIG.  5. Geometry for retro-reflection measurements.}
% \label{fidelityFig}
% \end{center}
% \end{figure}
% NOTE TO EDITOR: figure 5 goes near here.

Beam distortion is quantified by the fidelity factor introduced in 
Part I
\begin{equation}
F  \equiv  4
\frac{
\left|\int d^{3}x G({\bf x})H({\bf x})\right|^{2}
}{
\left[\int d^{3} x \left(|G({\bf x})|^{2} + |H({\bf x})|^{2}\right)\right]^{2}
}
\end{equation}
The greatest contrast occurs when $H$ is the 
phase-conjugate, or time-reverse of $G$, i.e., when $H({\bf{x}}) = 
G^{*}({\bf{x}})$.  In this situation $F = 1$.  
Under the approximation that the field envelopes obey the paraxial 
wave equations
\begin{eqnarray}
\frac{d}{dz} G & = & \frac{i}{2 k} \nabla_{\perp}^{2} G
\nonumber \\
\frac{d}{dz} H & = & \frac{-i}{2 k} \nabla_{\perp}^{2} H,
\end{eqnarray}
Green's theorem can be used to show that the volume integral is 
proportional to the mode-overlap integral
\begin{equation}
\int d^{3}x G({\bf{x}})H({\bf{x}}) = \Delta z \int dx dy G({\bf{x}})H({\bf{x}}),
\end{equation}
where the last integration is taken at any fixed $z$ and $\Delta z$ is 
the length of the interaction region.  Similarly, the beam powers are 
invariant under propagation and
the mode fidelity can be expressed entirely in terms of surface 
integrals as
\begin{eqnarray}
\label{PlanarModeFactor}
F  &= & 4\left|\int dxdy G({\bf{x}})H({\bf{x}})\right|^{2} 
\nonumber \\
& & 
\times
\left[\int dxdy \left(|G({\bf{x}})|^{2} + |H({\bf{x}})|^{2}\right)\right]^{-2}.
\end{eqnarray}
The overlap of $G$ and $H$ also determines the efficiency of coupling back 
into the fiber.  This allows us to determine $F$.  
In terms of $P_{\rm in}$, the power 
leaving the output fiber coupler and $P_{\rm ret}$, the power returned 
through the fiber after being retro-reflected, this is 
\begin{equation} 
F =
\frac{4}{(1+T^{2})^{2}} \frac{P_{\rm ret}}{\eta T P_{\rm in}}.  
\end{equation}
where $\eta = 0.883$ is the intensity transmission coefficient of the 
fiber and coupling lenses and $T = 0.92$ is the transmission 
coefficient for a single-pass through a cell window.  We find $F = 
0.36 \pm 0.03$.  The mode fidelity acts twice to reduce contrast, 
once as the drive beams enter the cell, and again on the photons 
leaving the cell.  This beam 
distortion has no effect on the incoherent scattering background, 
thus the visibility is reduced by $F^{2}$.

% \begin{figure}
% \begin{center}
% \epsfig{file=RbExpt.eps01,width=\mathsize,angle=0}
% \mycaption{FIG. 6.  Coincidence rates by photon-photon scattering theory: 
% adjusted for beam shape, finite detection time and detector area.}
% \label{PPExptFig}
% \end{center}
% \end{figure}
% NOTE TO EDITOR: figure 6 goes near here.

The finite time response of the detector system acts to disperse the 
coincidence signal over a larger time window.  This reduces the 
maximum contrast by a factor of 0.27 and increases the temporal width 
to 1.62 ns.  Similarly, the finite detector area reduces the maximum 
contrast by a factor of 0.81 and spreads the angular correlations by a 
small amount.  The resulting coincidence signal is shown in Fig. 
6.  Fitted to a Gaussian, the final signal contrast is 0.042 $\pm 
0.007$, where the uncertainty reflects the uncertainty in $F$.  This 
is consistent with the observed contrast of 0.044 $\pm$ 0.010.

\section{conclusion}

We have measured the temporal and angular correlations in 
photon-photon scattering mediated by atomic rubidium vapor.  We found 
good agreement between experiment and the perturbative theory 
presented in Part I. The observed temporal correlations are of the 
order of one nanosecond, much faster than the system can relax by 
radiative processes.  This is consistent with the prediction that the 
duration of the photon-photon interaction is determined by the 
inhomogeneous broadening of the vapor.

\bibliography{PPSB}

\begin{thebibliography}{1}

\bibitem{Mitchell=1999}
M.~W. Mitchell and R.~Y. Chiao, Submitted for publication  .

\bibitem{Imamoglu=1997}
A. Imamoglu, H. Schmidt, G. Woods, and M. Deutsch, Physical Review Letters {\bf
  79},  1467  (1997).

\bibitem{Grangier=1998}
P. Grangier, D. Walls, and K. Gheri, Physical Review Letters {\bf 81},  2833
  (1998).

\bibitem{Heidmann=1987}
A. Heidmann and S. Reynaud, Journal of Modern Optics {\bf 34},  923  (1987).

\bibitem{Gaeta=1988}
A. Gaeta and R. Boyd, Physical Review Letters {\bf 60},  2618  (1988).

\bibitem{Grangier=1986}
P. Grangier {\it et~al.}, Physical Review Letters {\bf 57},  687  (1986).

\bibitem{Pittman=1995}
T. Pittman, Y. Shih, D. Strekalov, and A. Sergienko, Phys. Rev. A {\bf 52},
  R3429  (1995).

\end{thebibliography}
\bibliographystyle{prsty}

\end{multicols}

\begin{figure}
\begin{center}
\epsfig{file=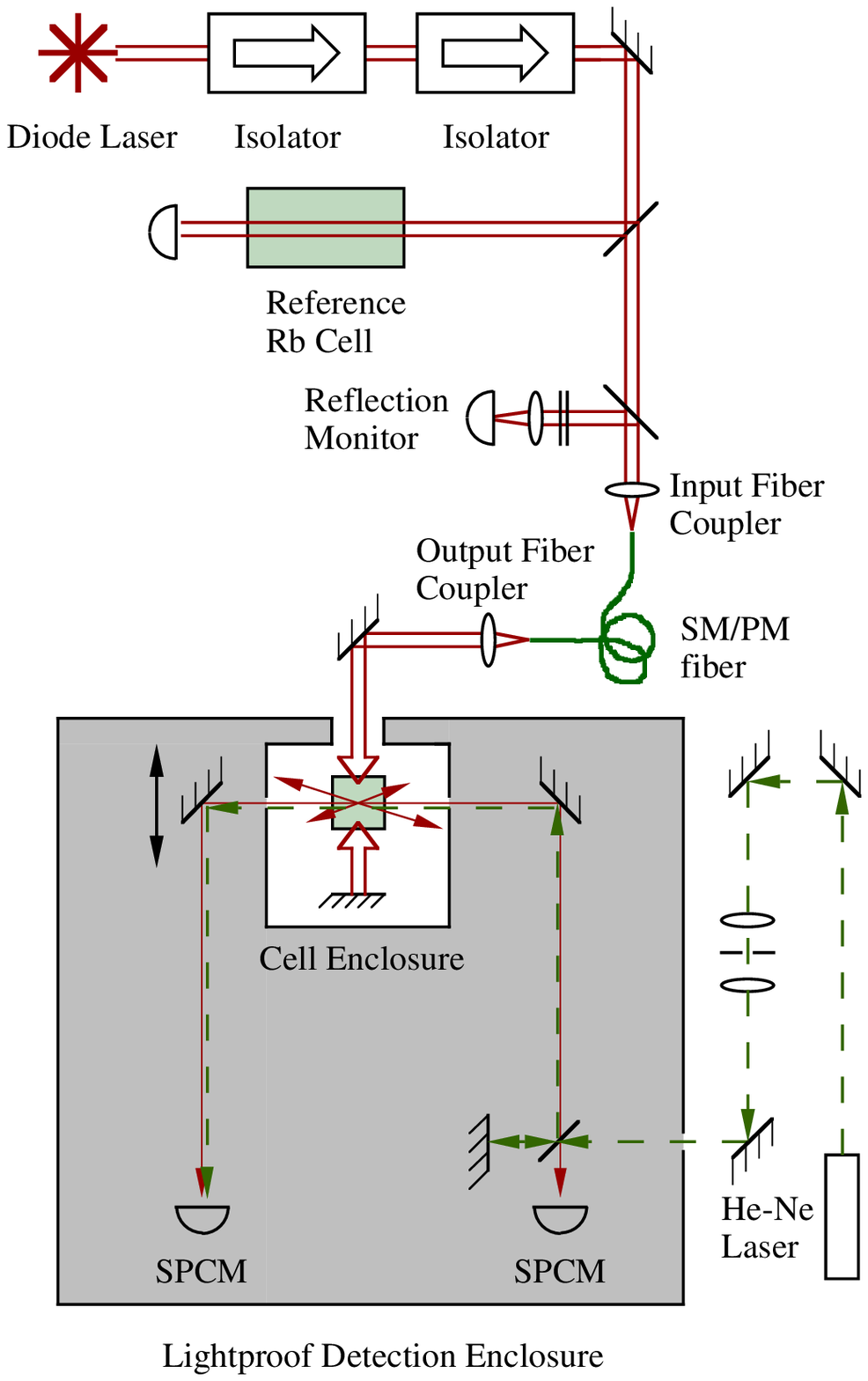,width = 2.5 in ,angle=0}
\caption{FIG. 1. Experiment schematic.}
\label{setupFig}
\end{center}
\end{figure}

\begin{figure}
\begin{center}
\epsfig{file=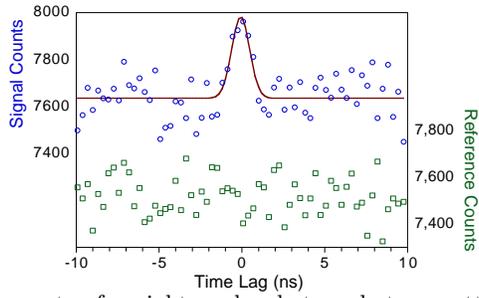,width = 2.5 in,angle=0}
\caption{FIG. 2.  Observed coincidence rates for right-angle photon-photon 
scattering.  Circles show data acquired with detectors aligned to 
collect back-to-back scattering products.  Squares show data acquired with 
detectors misaligned by 10 cm $\approx$ 0.14 radian.  The solid line 
is a Gaussian function fit to the data.  }
\label{ExptTimesFig}
\end{center}
\end{figure}

\begin{figure}
\begin{center}
\epsfig{file=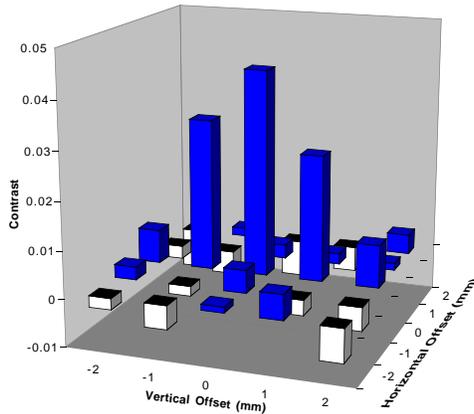,width = 2.5 in,angle=0} 
\caption{FIG. 3.  Signal contrast vs.  detector displacement.  A 
displacement of 1 mm corresponds to an angular deviation of 1.43 mrad.  
}
\label{AngularFig}
\end{center}
\end{figure}

\begin{figure}
\begin{center}
\epsfig{file=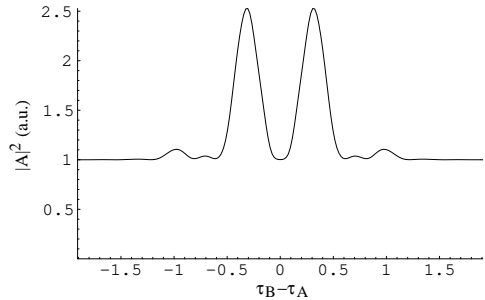,width=2.5 in,angle=0}
\caption{FIG. 4.  Coincidence rates by photon-photon scattering theory: 
ideal case.}
\label{PPIdealFig}
\end{center}
\end{figure}

\begin{figure}
\begin{center}
\epsfig{file=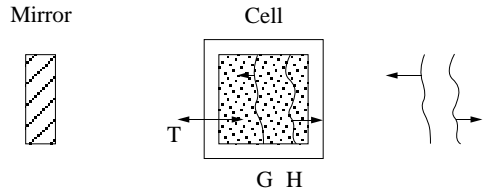,width = 2.5 in,angle=0}
\caption{FIG.  5. Geometry for retro-reflection measurements.}
\label{fidelityFig}
\end{center}
\end{figure}

\begin{figure}
\begin{center}
\epsfig{file=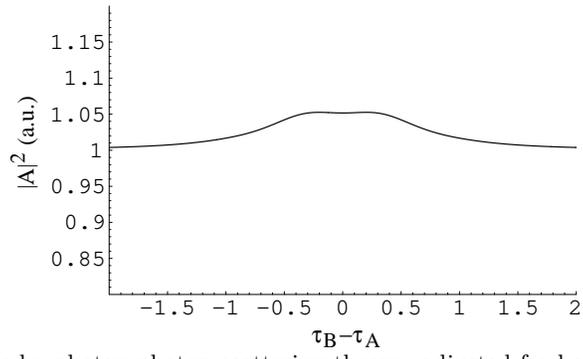,width=3.0 in,angle=0}
\caption{FIG. 6.  Coincidence rates by photon-photon scattering theory: 
adjusted for beam shape, finite detection time and detector area.}
\label{PPExptFig}
\end{center}
\end{figure}

\end{document}